\documentclass [aps, prb, amssymb, twocolumn, showpacs,superscriptaddress]{revtex4-1}
\usepackage{graphicx,epsfig, amsfonts, amssymb, amsmath}
\usepackage{bm}
\usepackage{times}
\usepackage{pifont}
\usepackage[colorlinks,citecolor=blue,linkcolor=red]{hyperref}
\usepackage{color}

\begin{document} 
 
\title{Nanoscale Spin Seebeck Rectifier: Controlling Thermal Spin Transport across Insulating Magnetic Junctions with Localized Spin}

\author {Jie Ren}\email{jieustc@gmail.com. Present address: Department of Chemistry, Massachusetts Institute of Technology, 77 Massachusetts Avenue, Cambridge, MA 02139, USA}
\affiliation{Theoretical Division, Los Alamos National Laboratory, Los Alamos, New Mexico 87545} 
\author{Jonas Fransson}
\affiliation{Department of Physics and Astronomy, Uppsala University, Box 516, SE-751 21 Uppsala, Sweden}
\author{Jian-Xin Zhu}
\affiliation{Theoretical Division, Los Alamos National Laboratory, Los Alamos, New Mexico 87545} 
\affiliation{Center for Integrated Nanotechnologies, Los Alamos National Laboratory, Los Alamos, New Mexico 87545, USA}

\date{\today}

\begin{abstract}
The spin Seebeck effect is studied across a charge insulating magnetic junction, in which thermal-spin conjugate transport is assisted by the exchange interactions between the localized spin in the center and electrons in metallic leads. 
We show that, in contrast with bulk spin Seebeck effect, the figure of merit of such nanoscale thermal-spin conversion can be infinite, leading to the ideal Carnot efficiency in the linear response regime. We also find that in the nonlinear spin Seebeck transport regime, the device possesses the asymmetric and negative differential spin Seebeck effects. 
{  In the last, the situations with leaking electron tunneling are also discussed.}
This nanoscale thermal spin rectifier, by tuning the junction parameters, can act as a spin Seebeck diode, spin Seebeck transistor and spin Seebeck switch, which could have substantial implications for flexible thermal and information control in molecular spin caloritronics.
\end{abstract}



\maketitle
\section{Introduction}
Energy waste is a severe bottleneck in the supply of sustainable energy to any modern economy.
Besides developing new energy sources, the global energy crisis can be alleviated by re-utilizing the wasted energy. In view of the fact that about 90\% of the world's energy utilization occurs in the form of heat, effective heat control and conversion become critical~\cite{Chu}.
To meet the desire, phononics~\cite{phononics} has been proposed to control heat energy and information in a similar style as controlling electric current and signal in electronics. Various functional thermal devices such as thermal rectifiers and transistors are then designed, essentially based on two intriguing properties: the heat diode effect and negative differential thermal conductance [e.g., see Refs.~\onlinecite{phononics, Ren2013PRB87}].

Meanwhile, the investigation on interplay of spin and heat transport has attracted great interest.
In particular, spin Seebeck effect has been widely observed recently~\cite{Uchida2008Nature, Jaworski2010NatureMat, Breton2011Nature, Uchida2010NatureMat, Uchida2010APL, Kikkawa2013PRL, Qu2013PRL, Jaworski2012Nature}; that is, the temperature bias can produce a pure spin current {\it in the absence of} electron current. Since then, the spin Seebeck effect has ignited a upsurge of renewed research interest, because it acts as a new method of functional use of waste heat as spin caloritronics~\cite{BauerReview} and opens more possibilities for spintronics~\cite{spintronics} and magnonics~\cite{magnonics}, which allows us to realize non-dissipative information and energy transfer without Joule heating~\cite{Kajiwara2010Nature, TMI} and to construct thermoelectric devices upon new principles \cite{Kirihara2012NatureMat}.

By integrating the spin Seebeck effect with concepts from phononics~\cite{phononics,Ren2013PRB87}, the \emph{asymmetric spin Seebeck effect} (ASSE) has recently been discovered both in metal/insulating magnet interfaces~\cite{RenSSE1} and magnon tunneling junctions~\cite{RenSSE2}, which leads us to spin Seebeck diodes to rectify the thermal energy and spin information~\cite{RenSSE1,RenSSE2}. Similar rectification of spin Seebeck effect is also discussed in other insulating magnetic systems~\cite{Juzar,add}.
Beyond spin Seebeck diodes, the \emph{negative differential spin Seebeck effect} (NDSSE) has been further uncovered both in metal/insulating magnet interfaces~\cite{RenSSE1} and magnon tunneling junctions~\cite{RenSSE2}, i.e., increasing thermal bias gives the decreasing spin current. This NDSSE is crucial to realize spin Seebeck transistors~\cite{RenSSE1, RenSSE2}.

Developing nanoscale spin Seebeck devices with such ASSE and NDSSE is a great challenge not only for fundamental science but also for practical applications. By utilizing and controlling spin Seebeck effects at atomic/molecular levels that could benefit from the scalability and tunability of nanodevices, we may transform the field of molecular spintronics~\cite{MS1,MS2,MS3,MS4} to the possible ``molecular spin caloritronics'', where we can have flexible control of spin-mediated energy flow or thermal-mediated spin current. Such nanoscale spin caloritronics would have potential impact on a variety of new technologies but still requires a better understanding of spin Seebeck effects in the test bed of nanoscale junctions.

\begin{figure}
\scalebox{0.38}[0.38]{\includegraphics{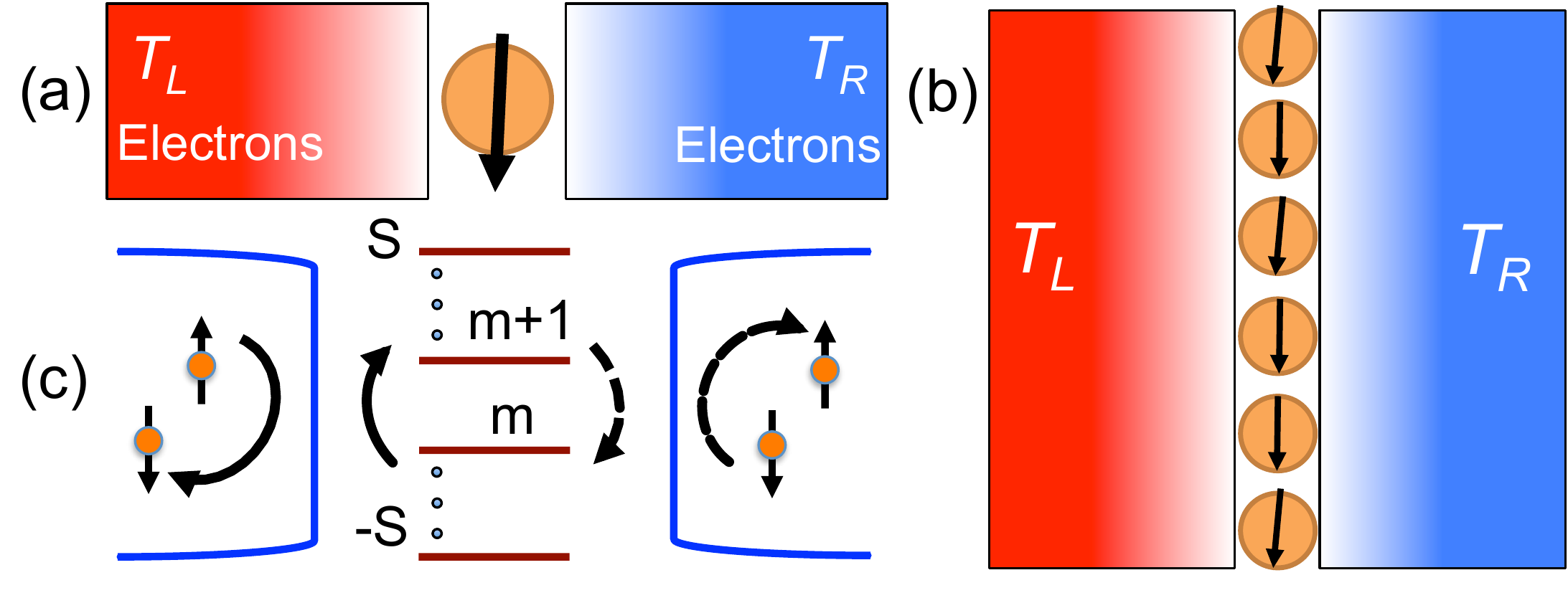}}
\vspace{-4mm}   
\caption{(color online) (a) Scheme of the charge insulating magnetic junction with a localized spin. (b) Multiple parallel transport channels, which can be constructed to enhance the transfer signal. (c) Schematic illustration of the dynamics without magnetic anisotropy: The spin transfer is assisted by the excitation and relaxation of the central spin, as indicated by the arrows. 
} 
\label{fig1}
\end{figure}

In this work, we study the nonequilibrium spin Seebeck transport through an insulating magnetic molecular quantum dot, which is in contact with ferromagnetic leads held at different temperatures. 
We show that in contrast with bulk spin Seebeck effect,  the thermal-spin conversion in such a molecular spin caloritronic device can reach an infinite figure of merit, which indicates the ideal Carnot efficiency. In the nonlinear spin Seebeck transport regime, we also find that the device exhibits the ASSE and NDSSE. 
In the last, the situations with leaking electron tunneling are also discussed.
This nanoscale thermal spin rectifier, by tuning the junction parameters, can act as a spin Seebeck diode, spin Seebeck transistor, and spin Seebeck switch, which we believe could have substantial implications for flexible thermal and spin information control in molecular spin caloritronics.

\section{Model and Method} 
We consider {  a phenomenological model that describes} a localized effective spin coupled with two metallic ferromagnetic leads [see Fig.~\ref{fig1}(a)], with the total Hamiltonian 
\begin{equation}
H=\sum_{v=L,R}H_v+\sum_{v=L,R}V_v+H_s.
\label{eq:H}
\end{equation}
{  The central local spin may represent an insulating molecule magnet \cite{Schnack} or a ferromagnetic nanoparticle \cite{Wiesen} found in a nanoscale single-domain state and thus is described by the effective macrospin~\cite{macrospin}, as:} 
\begin{equation}
H_s=\omega_0S^z+D(S^z)^2,
\label{eq:macrospin}
\end{equation}
with $D$ the easy-axis anisotropy and $\omega_0$ the intrinsic energy of the local spin controlled by external fields or proximity effects. 
Two metallic ferromagnetic leads are severally in equilibrium with temperatures $T_v$ and are described by the Stoner-model Hamiltonians: 
\begin{equation}
H_{v}=\sum_{\sigma{k}\in{v}}\epsilon_{k\sigma}c^{\dag}_{k\sigma}c_{k\sigma},
\end{equation}
where $c^{\dag}_{k\sigma}~(c_{k\sigma})$ denotes the creation (annihilation) operator of electrons with momentum $k$, spin $\sigma$, and energy $\epsilon_{k\sigma}$ that may have different spin-resolved density of states (DOS).  
The spin-lead coupling is described by the { local} exchange interaction 
\begin{equation}
V_{v}=J_v\sum_{\sigma k,\sigma'k'\in v}c^{\dag}_{k\sigma}\bm{\tau}_{\sigma\sigma'}c_{k'\sigma'}\cdot\bm{S},
\label{eq:coupling}
\end{equation}
which couples the central local spin $\bm{S}$ to the electronic spin $c^{\dag}_{k\sigma}\bm{\tau}_{\sigma\sigma'}c_{k'\sigma'}$ in the lead. $J_v$ denotes the exchange coupling strength to the $v$th lead and $\bm{\tau}_{\sigma\sigma'}$ is the Pauli matrix elements. 

{ We note that the direct electron tunneling and spin exchange between two metallic leads are neglected. The reason is that the coupling of electrons in two leads originates from the wave-function overlap, which generally decays exponentially with the distance~\cite{Cris}. The distance between two leads doubles the central spin-lead distance, which could make the lead-lead interaction a few orders of magnitude smaller compared to the central spin-lead interaction, thus negligible. A recent work shows that, however, the cotunneling that survives only at the extremely low temperature~\cite{NatureNano} can support the long-range tunneling. But since the spin Seebeck transport at high (room) temperatures is of our prime interest, the cotunneling is not included in our present scheme. Here we also neglect the possible spin-assisted electron tunneling between the two leads. The discussion on the effect of spin-assisted electron tunneling is deferred to Sec. \ref{discussion}.}

{ 
Our setup, reminiscent of the junction of metal/insulating magnet/metal~\cite{Kajiwara2010Nature},  can be regarded as two copies of the metal/insulating magnet interface studied in Ref.~\cite{Cris}, where the insulating layer has a large electron band gap so that only exchange interactions at boundaries with two leads are responsible for the spin Seebeck transport without electric current.
In fact,  the signal transmission through a sandwiched metal/magnetic insulator/metal junction has already been observed in experiment~\cite{Kajiwara2010Nature}. Our system of a single local spin in the insulating central part is considered as the minimal phenomenological model to mimic the sandwich setup, because in ferromagnets at nanoscale spins are tightly coupled and form an effective coarse-grained macrospin, as described by Eq.~(\ref{eq:macrospin})~\cite{macrospin}.
}

{ 
Note our this scheme is different from the earlier setups in Refs.~\onlinecite{JXZhu,Elste,Wang}, where electron tunneling transports are considered and the exchange coupling is between the tunneling electron spin and the local molecular spin.
The important observation in the present work is that even in the charge insulating case, where electron transfer is essentially quenched across the junction, we can still have the spin Seebeck transport assisted by the exchange coupling $V_v$,} which precisely speaking has three contributions: 
\begin{equation}
V_{v}\!=\!J_v\!\!\!\sum_{k,k'\in{v}}[S^{z}(c^{\dag}_{k\uparrow}c_{k'\uparrow}-c^{\dag}_{k\downarrow}c_{k'\downarrow})+S^{-}c^{\dag}_{k\uparrow}c_{k'\downarrow}+S^{+}c^{\dag}_{k\downarrow}c_{k'\uparrow}],
\label{eq:coupling2}
\end{equation}
with $S^{\pm}$ the spin-raising (-lowering) operators.
The first term renormalizes the intrinsic energy $\omega_0$ of the central spin into a new one $\Omega_0$ though the proximity effect. 
Only the last two terms are responsible for the spin Seebeck transport. In practice, multiple macrospins hosted by molecular quantum dots can be constructed in between two leads, as illustrated in Fig.~\ref{fig1}(b), to form parallel transport channels so as to enhance the transfer power and signal.

We start with the Liouville-von Neumann equation for the reduced density matrix of the central spin, which is represented in the eigenstate basis $|m\rangle$ of $S^z$. For the weak system-lead coupling and Markovian limit, the dynamics of the central spin is obtained as a Pauli master equation~\cite{Timm,Misiorny1,Lu,Wang,Elste}:
\begin{equation}
\frac{\partial{P}_{m}}{\partial{t}}=\sum_{n}\left(P_{n}k_{n\rightarrow{m}}-P_{m}k_{m\rightarrow{n}}\right),
\label{eq:ME}
\end{equation}
with $P_m$ the probability of the spin state $|m\rangle$. The transition rate has two contributions from the left and right leads: $k_{m\rightarrow{n}}=k^L_{m\rightarrow{n}}+k^R_{m\rightarrow{n}}$, with the initial and final states constrained by nearest state transitions $n=m\pm{1}$. The spin current from the central system into the right can be derived from the Heisenberg equation $I_s:=\langle\frac{d}{dt}\sum_{\sigma\sigma'k\in{R}}c^{\dag}_{k\sigma}\frac{\tau^z_{\sigma\sigma'}}{2}c_{k\sigma'}\rangle=\frac{i}{\hbar}J_R\sum_{k,k'\in{R}}\langle{S}^+c^{\dag}_{k\downarrow}c_{k'\uparrow}-S^-c^{\dag}_{k\uparrow}c_{k'\downarrow}\rangle$, yielding
\begin{equation}
I_s=\sum_{m}\left(P^{\text{ss}}_{m+1} k^R_{m+1\rightarrow{m}}-P^{\text{ss}}_{m}k^R_{m\rightarrow{m+1}}\right),
\label{eq:Is}
\end{equation} 
where the steady state probability $P^{\text{ss}}_m$ is calculated by setting Eq.~(\ref{eq:ME}) equal to zero. The spin current at left can be obtained equally.
Straightforward calculations [similar to \onlinecite{Timm,Misiorny1,Lu,Wang,Elste}] lead to the rate expressions:
\begin{eqnarray}
k^v_{m\rightarrow{m\pm1}}&=&\frac{2{\pi}J^2_v}{\hbar}\left|\langle{m\pm1}|S^{\pm}|m\rangle\right|^2\mathcal{W}^{\pm},
\label{eq:rate1}
\end{eqnarray}
with
\begin{eqnarray}
\mathcal{W}^{\pm}&=&\int^{\infty}_{-\infty}d\epsilon\rho_{v\uparrow}(\epsilon+\Omega_0+[2m\pm1]D)\rho_{v\downarrow}(\epsilon)   \nonumber\\
&\times&f^{\pm}_{v\uparrow}(\epsilon+\Omega_0+[2m\pm1]D)f^{\mp}_{v\downarrow}(\epsilon),
\label{eq:rate11}
\end{eqnarray}
where $\left|\langle{m\pm1}|S^{\pm}|m\rangle\right|^2=S(S+1)-m(m\pm1)$ with $S$ the spin length; $\rho_{v\sigma}(x)$ denotes the DOS for electrons with spin $\sigma$ and energy $x$ in the lead $v=(L,R)$; $f^+_{v\sigma}(x):=1-f^-_{v\sigma}:=[e^{(x-\mu_{v\sigma})/(k_BT_v)}+1]^{-1}$ is the Fermi-Dirac distribution in the lead $v$ with spin-dependent chemical potential $\mu_{v\sigma}$ at temperature $T_v$. 

The rates have clear physical meanings [see Fig.~\ref{fig1}(c)]: $k^v_{m+1\rightarrow{m}}{\propto}f^{-}_{v\uparrow}(\epsilon+\varepsilon_m)f^{+}_{v\downarrow}(\epsilon)$, in which $\varepsilon_m:=\Omega_0+[2m+1]D$, depicts the scattering rate of a spin-down electron in lead $v$ at energy $\epsilon$ into a spin-up state in the same lead at energy $\epsilon+\varepsilon_m$, accompanied by reducing the central spin state from $m+1$ to $m$.  $k^v_{m\rightarrow{m+1}}{\propto}f^{+}_{v\uparrow}(\epsilon+\varepsilon_m)f^{-}_{v\downarrow}(\epsilon)$ describes the scattering rate of a spin-up electron in lead $v$ at energy $\epsilon+\varepsilon_m$ into a spin-down state in the same lead at energy $\epsilon$, accompanied by increasing the central spin state from $m$ to $m+1$. These exchange transitions conserve the spin angular momentum and are responsible for the pure spin transfer as depicted in Fig.~\ref{fig1}(c). They also satisfy the detailed-balance-like relation: $k^v_{m+1\rightarrow{m}}/k^v_{m\rightarrow{m+1}}=\exp{(\frac{\varepsilon_m-{\delta}\mu^s_v}{k_BT_v})}$ with $\delta{\mu}^s_v:=\mu_{v\uparrow}-\mu_{v\downarrow}$ denoting the spin accumulation in the lead $v$. To make this clear, we can rewrite Eqs.~(\ref{eq:rate1}, \ref{eq:rate11}) as
\begin{equation}
k^v_{m\rightleftharpoons{m+1}}=\pm\frac{2{\pi}J^2_v}{\hbar}\left|\langle{m+1}|S^{+}|m\rangle\right|^2N_{v}(\pm\varepsilon_m)C_{v}(\varepsilon_m), 
\label{eq:rate2}   
\end{equation}
where 
\begin{equation}
C_{v}(\varepsilon_m)=\int^{\infty}_{-\infty}d\epsilon\rho_{v\uparrow}(\epsilon+\varepsilon_m)\rho_{v\downarrow}(\epsilon)[f^{+}_{v\downarrow}(\epsilon)-f^{+}_{v\uparrow}(\epsilon+\varepsilon_m)]
\label{eq:C}
\end{equation}
 is an integral generally depending on the energy $\varepsilon_m$, 
 the chemical potentials, the temperatures and the overlap between two spin-resolved DOS; $N_{v}(\pm\varepsilon_m)=[\exp{(\pm\frac{\varepsilon_m-{\delta}\mu^s_v}{k_BT_v})}-1]^{-1}$ are Bose-Einstein distributions with the ratio $-N_{v}(-\varepsilon_m)/N_{v}(\varepsilon_m)=\exp{(\frac{\Omega_0+[2m+1]D-{\delta}\mu^s_v}{k_BT_v})}$. { Note, from Eqs.~(\ref{eq:rate1}, \ref{eq:rate11}) to Eqs.~(\ref{eq:rate2}, \ref{eq:C}), we have utilized the equalities: $f_{v\uparrow}(\epsilon+\varepsilon_m)[1-f_{v\downarrow}(\varepsilon)]\equiv N_v(\varepsilon_m)[f_{v\downarrow}(\epsilon)-f_{v\uparrow}(\epsilon+\varepsilon_m)]$ and $[1-f_{v\uparrow}(\epsilon+\varepsilon_m)]f_{v\downarrow}(\epsilon)\equiv [1+N_v(\varepsilon_m)][f_{v\downarrow}(\epsilon)-f_{v\uparrow}(\epsilon+\varepsilon_m)]$.
 }

\section{Results}  
Without loss of generality we focus on the $S=1/2$ case, where the anisotropy is irrelevant, {although our above formulations are valid for general situations.} For large spin cases, we find the large anisotropy can inverse the sign of thermal-spin transport due to the fact that large $D$ can invert the spin eigen-levels $m\Omega_0+m^2D$ to make them parabolic instead of linearly equal-spaced at $D=0$. However, usually the anisotropy $D\sim\mu${eV} is much smaller than other energy scales ($\sim${meV}) of interest. Thus, the effect of magnetic anisotropy will be insignificant except in the extremely low temperature regime, where the Kondo effect may play a role~\cite{Elste,Liang,Park,Otte,Romeike,Wegewijs,Leuenberger,Gonzlez,WangRQ,Cornaglia} and is beyond the scope of the present work.  For large spin cases without anisotropy, they share quantitatively the same behaviors as the spin-half case that we will discuss in detail in the following. 

{ 
For the $S=1/2$ case, the spin current can be analytically obtained from Eqs.~(\ref{eq:Is}) and~(\ref{eq:rate2}), as:
\begin{align}
I_s&= \frac{k^L_{01}k^R_{10}-k^R_{01}k^L_{10}}{k^L_{01}+k^L_{10}+k^R_{10}+k^R_{01}}
\nonumber\\
&=\frac{\frac{2\pi}{\hbar}J^2_LJ^2_RC_L(\Omega_0)C_R(\Omega_0)[N_L(\Omega_0)-N_R(\Omega_0)]}{J^2_LC_L(\Omega_0)[1+2N_L(\Omega_0)]+J^2_RC_R(\Omega_0)[1+2N_R(\Omega_0)]},
\label{eq:TLflux}
\end{align}
where we use $k^{v}_{01}, k^{v}_{10}$ to simplify the notations $k^{v}_{-\frac{1}{2}\rightarrow \frac{1}{2}},k^{v}_{\frac{1}{2}\rightarrow -\frac{1}{2}}$ by re-expressing the spin up and down states with state $1$ and $0$.}
Clearly, the pure spin transfer is driven by the difference $[N_L(\Omega_0)-N_R(\Omega_0)]$, from which we can see that, merely the spin accumulation difference (spin voltage) $\Delta\mu_s=\delta\mu^s_L-\delta\mu^s_R\neq0$ or the temperature bias $\Delta{T}=T_L-T_R\neq0$ is able to generate nonzero spin current, while the chemical potential difference between two leads $\mu_L\neq\mu_R$ can not. This emphasizes that the spin Seebeck transport here is not driven by the electric bias, but by the thermal bias or spin (voltage) bias. The thermal transport can also be similarly formulated and we obtain the heat current as $I_Q=\Omega_0I_s$.

\subsection{Linear transport properties}
Let us first examine the spin thermal transport coefficients in the linear response regime.  Considering $\delta\mu^s_{L,R}=\pm\Delta\mu_s/2, T_{L,R}=T\pm\Delta T/2$, we are able to expand the spin and heat currents to the first order of spin voltage and thermal bias ($\Delta\mu_s, \Delta{T}\rightarrow0$)~\cite{Callen,Mahan}, yielding
\begin{eqnarray}
\binom{I_s}{I_Q}=
\left(
\begin{array}{cc}
\mathcal L_0  &  \mathcal L_1  \\
 \mathcal L_1   &  \mathcal L_2
\end{array}
\right)
\binom{\Delta \mu_s}{\Delta T/T},
\label{eq:linear}
\end{eqnarray}
where 
\begin{equation}
\mathcal L_n=\Omega_0^n\frac{2\pi}{\hbar}\frac{J^2_LJ^2_RC_LC_R}{2T(J^2_LC_L+J^2_RC_R)\sinh[\Omega_0/T]} ,
\end{equation}
with $C_{L,R}$ at zero bias $\Delta\mu_s, \Delta T=0$.
Clearly, $\mathcal G=\mathcal L_0$ denotes the spin conductance for the pure spin transfer generated by the spin voltage $\Delta\mu_s$; $\mathcal{S}_s:=-\Delta \mu_s/\Delta{T}|_{I_s=0}=\mathcal L_1/(\mathcal L_0T)=\Omega_0/T$ is the spin Seebeck coefficient, depicting the power of generating spin voltage by the temperature bias; $\Pi:=I_Q/I_s|_{\Delta{T}=0}=\mathcal L_1/\mathcal L_0=\Omega_0$ is the spin Peltier coefficient, depicting the power of heating or cooling carried by per unit spin current. One can see that the Kelvin relation~\cite{Callen} (one sort of Onsager reciprocal relations) $\Pi=\mathcal{S}_sT$ is fulfilled. Clearly, only the spin transport and  the thermal transport are conjugated to each other. The thermal bias is able to generate the spin current in the absence of electron transport so that it is a pure spin Seebeck effect; The spin voltage is able to generate the heat current without electric current so that it is a pure spin Peltier effect. This situation is different from previous thermal spin transport studies where electronic current and voltage are involved \cite{Dubi,Wang,Cornaglia,Johnson,Takezoe}. 

The thermal-spin conversion efficiency is given by $\eta=\eta_c\frac{\sqrt{ZT+1}-1}{\sqrt{ZT+1}+1}$, where $\eta_c$ is the ideal Carnot efficiency and the figure of merit is $ZT:=\mathcal{G}\mathcal{S}^2_sT/\kappa_s$~\cite{Callen}. Here, needs to be pointed out is that in the denominator the thermal conductance is defined at zero spin current $\kappa_s:=I_Q/\Delta{T}|_{I_s=0}$~\cite{Ren2012PRB85, Mahan}, not at zero spin bias $\Delta\mu_s=0$. In other words, $\kappa_s$ should be correctly obtained as $\kappa_s:=(\mathcal L_2-\mathcal L^2_1/\mathcal L_0)/T$, not as $I_Q/\Delta T|_{\Delta\mu_s=0}=\mathcal L_2/T$.~\cite{Ren2012PRB85, Mahan} Interestingly, since Eq.~(\ref{eq:linear}) has the proportionality between the heat and spin currents: $I_Q=\Omega_0I_s$, one will get zero heat current at zero spin current, which leads to  $\kappa_s=0$ so that $ZT\rightarrow\infty$. This infinite figure of merit $ZT$ is not unphysical. It  just tells us the efficiency of the thermal-spin conversion approaches to the ideal Carnot efficiency of the device and is still upper-bounded by $1$. The ideal Carnot efficiency resulting from the strict proportionality between the spin and heat currents, was similarly discussed in other contents of energy conversions~\cite{Kedem, Humphrey, Broeck, Keiji1}, called the thermodynamic tight-coupling limit.

{  Note that this tight-coupling induced infinite $ZT$ originates from the strict proportionality between the spin and heat currents, which is valid for the ideal case  without the magnetic anisotropy ($D=0$). In reality, finite anisotropy will distort the linear equal-spaced spin levels, which in turn removes the strict proportionality between the heat and spin currents. Moreover, the ignored electron transfer, as well as the photon-carried radiation heat transfer in reality (possibly with phononic thermal transfer~\cite{phononics}), will contribute finite thermal conductances to the denominator of $ZT$, as $ZT:=\mathcal{G}\mathcal{S}^2_sT/(\kappa_s+\kappa_e+\kappa_{ph})$, so that the infinity will be removed although $ZT$ may still be large. 
Last but not least, one should be aware of the fact that $ZT$ is a linear response quantity that merely characterizes the performance  close to zero power and has little meaning outside the linear response regime.
Even infinite $ZT$ does not give the best performance at finite power (for example, see Ref.~\cite{Whitney} and references therein), which depends also on the short-circuit spin current, the maximum output power, and the filling factor of the system. }

\subsection{Nonlinear transport properties}
In what follows, we focus on the spin Seebeck effect in the nonlinear response transport regime. 
We fix zero spin voltage $\mu_{v\uparrow}=\mu_{v\downarrow}=\mu_v$ for both leads and only consider the thermal-generated spin current with temperature bias. Controlling such thermal spin transport can be achieved by either tuning $J_v$ or $C_v$, the latter of which depends on chemical potentials, temperatures and  spin-resolved DOS overlaps [see Eq.~(\ref{eq:C})], thus offering us plenty of intriguing spin Seebeck properties. Without loss of generality, we assume the leads are confined in two dimension as thin films that have been used for the longitudinal spin Seebeck measurement~\cite{Kikkawa2013PRL}. Thus, for the up-polarized lead we can set the DOS as $\rho_{v\uparrow}(\epsilon)=\rho^0_{v\uparrow}\Theta(\epsilon), \rho_{v\downarrow}(\epsilon)=\rho^0_{v\downarrow}\Theta(\epsilon-\Delta_v)$ while for the down-polarized lead the DOS are $\rho_{v\uparrow}(\epsilon)=\rho^0_{v\uparrow}\Theta(\epsilon-\Delta_v), \rho_{v\downarrow}(\epsilon)=\rho^0_{v\downarrow}\Theta(\epsilon)$. 

\begin{figure}
\scalebox{0.44}[0.44]{\includegraphics{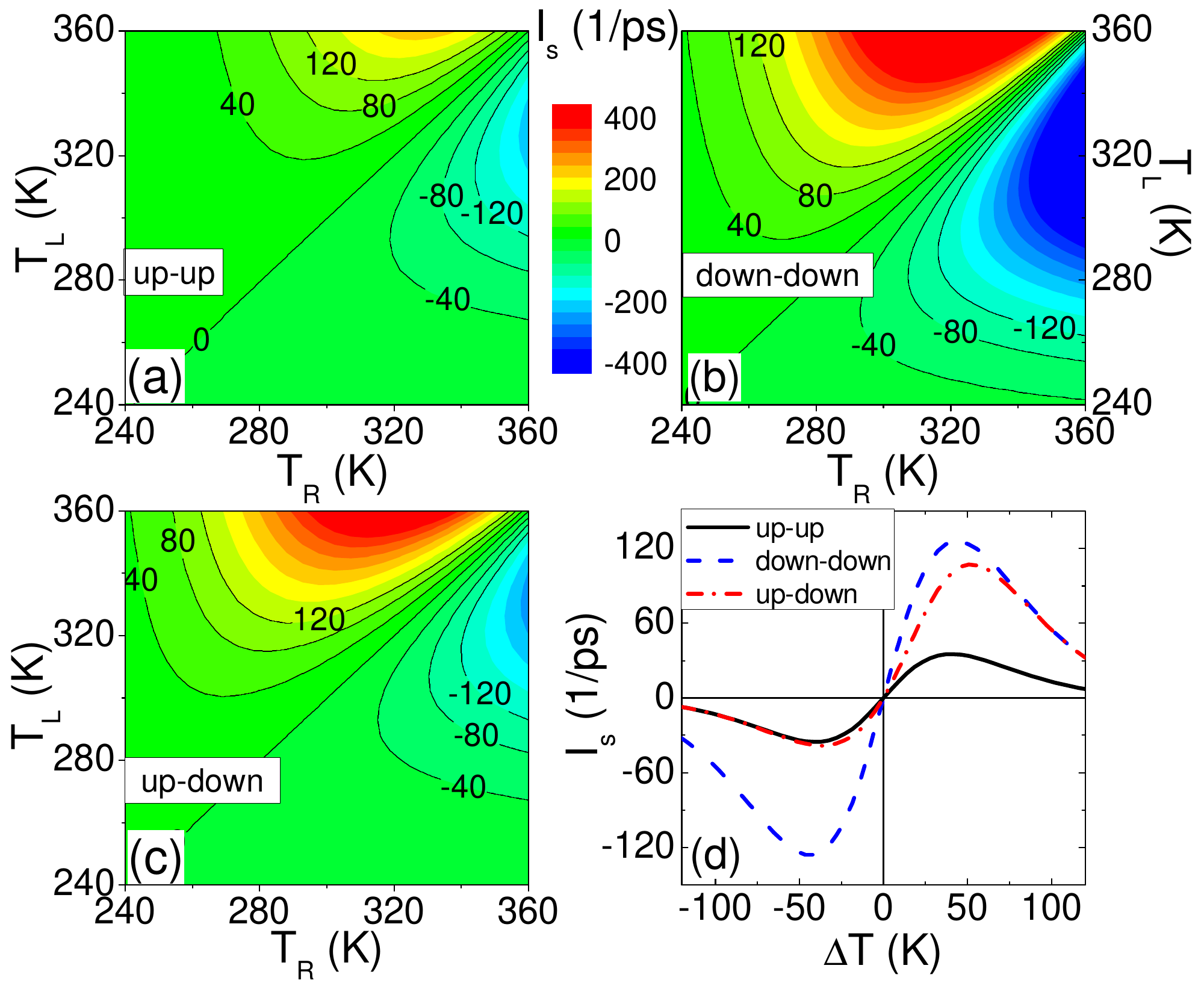}}
\vspace{-8mm}   
\caption{(color online). Tunable thermal spin rectifier with different lead polarizations: (a) up-up; (b) down-down; (c) up-down. 
Parameters are $\Delta_v=0.5$ eV, $\rho^0_{v\uparrow}=\rho^0_{v\downarrow}=0.4$/$\mu$eV, $\mu_{L,R}=0.15$ eV, $J_{L,R}=10$ meV, $\Omega_0=30$ meV. For (d), $T_{L,R}=T\pm\Delta{T}/2$ with $T=300$ K. The asymmetric lead polarization offers the spin Seebeck diode action. } 
\label{fig2}
\end{figure}

\subsubsection{Spin Seebeck diode.} 
Figure~\ref{fig2} shows that one can tune the spin Seebeck transport by changing the spin polarizations of two ferromagnetic leads. It is known that thermoelectric effects depend on the magnetic configurations of ferromagnetic leads~\cite{Krawiec,Swirkowicz,Sothmann2012EPL}. But the spin Seebeck effect is distinct in the sense that electron transport is absent. When the fully-polarized directions of two leads are tuned from up-up to down-down, the spin Seebeck transport is dramatically enhanced [see Fig.~\ref{fig2}(a), (b) and (d)]. This is because the DOS overlap $\rho_{v\uparrow}(\epsilon+\Omega_0)\rho_{v\downarrow}(\epsilon)$ of the down polarization case is larger than that of the up case. The increased DOS overlap increases the effective system-lead coupling $({\propto}J^2_vC_v)$ as indicated in Eq.~(\ref{eq:C}), which in turn increases the thermal spin current Eq.~(\ref{eq:TLflux}). If two leads have opposite polarizations [see Fig.~\ref{fig2}(c) and (d) for the case of left being spin-up and right spin-down], we can even have an asymmetric $I_s$ with respect to the thermal bias $\Delta{T}=T_L-T_R$, a rectification of spin Seebeck effect. In other words, we obtain the ASSE and a {spin Seebeck diode}, which acts as a good thermal spin conductor in one direction but acts as a poor spin Seebeck conductor or even an insulator in the opposite direction \cite{RenSSE1,RenSSE2,Juzar,add}. This is due to the fact that when two leads have different polarizations, the different spin-resolved DOS overlaps in the integral of $C_v$ make $C_L$ and $C_R$ have different responses to temperature change. If we keep one lead as ferromagnetic metal but set the other one as normal metal, we will have the similar ASSE.

\subsubsection{Spin Seebeck transistor.} 
Moreover, Fig.~\ref{fig2} shows that for different fully-polarized lead configurations, although at small temperature bias we have linear increasing of spin current, at large bias we generally have the phenomenon of NDSSE, i.e., increasing thermal bias anomalously decreases the spin Seebeck current~\cite{RenSSE1}, which is essential for constructing the spin Seebeck transistor~\cite{RenSSE2}. This negative differential transport is obtained due to the suppression of thermal-excited coexistence of electrons with both spins. The spin transport requires that electrons are scattered between spin-up and spin-down states. Although increasing one lead temperature and decreasing the other one will increase the thermal bias that subsequently increases $|N_L(\Omega_0)-N_R(\Omega_0)|$, the lowering temperature of one lead will severely suppress the thermal-excited minority electronic spin. As a consequence, the effective coupling $({\propto}J^2_vC_v)$ between the central spin and the electronic spins in the cold lead decreases, through the decreasing integral $C_v$ when decreasing temperature. Once the effective coupling decreases faster than the increasing of the thermal bias, negative differential spin Seebeck effect emerges.

\begin{figure}
\scalebox{0.38}[0.38]{\includegraphics{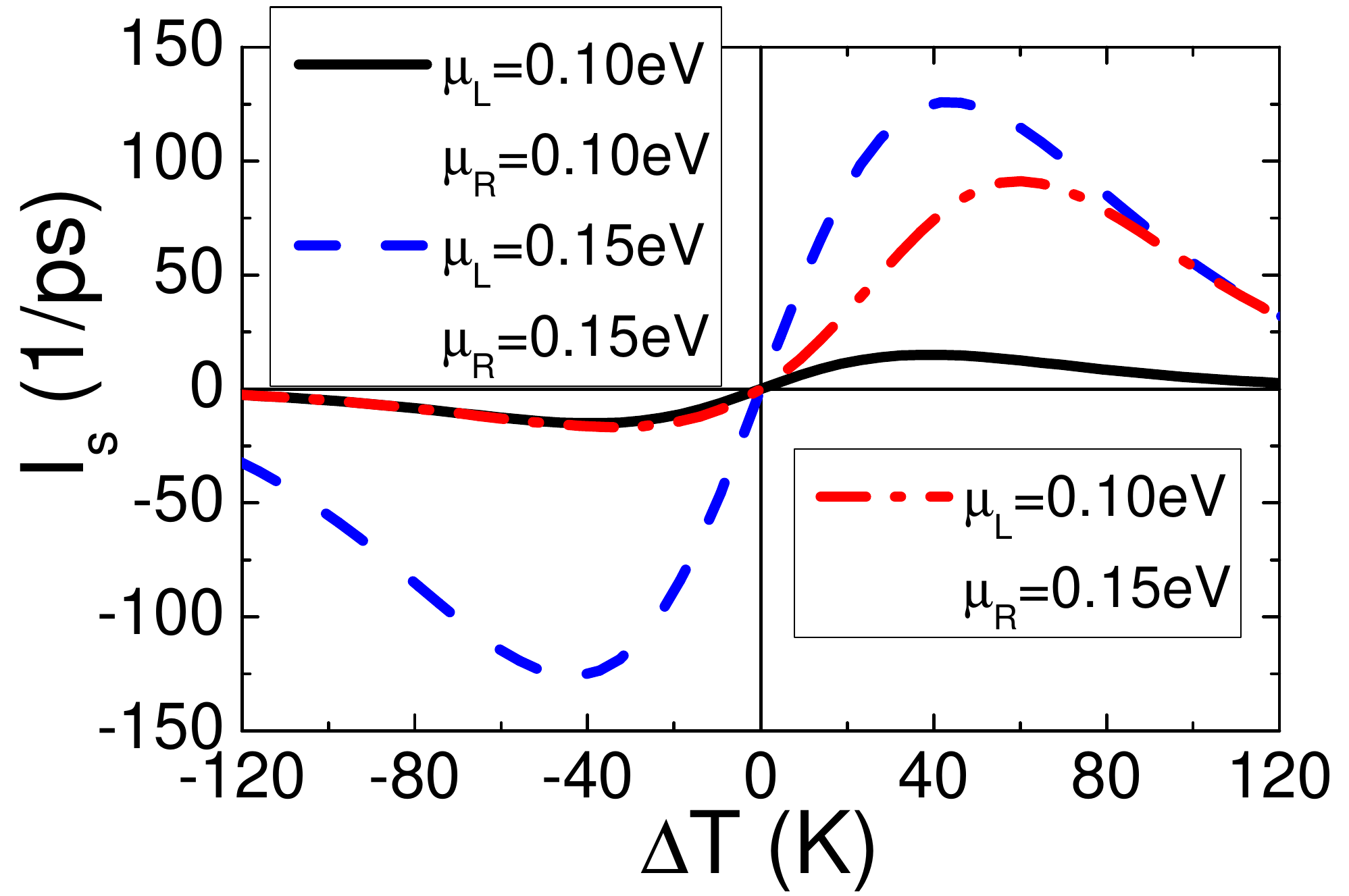}}
\vspace{-5mm}   
\caption{(color online). Tunable thermal spin switch via varying chemical potentials by applied electric field. Two leads are both down-polarized. Parameters not specified are the same as in Fig.~\ref{fig2}(d). Clearly, the chemical potential difference, although it acts as a control factor of the asymmetric spin Seebeck effect, can not generate the thermal spin transport. } 
\label{fig3}
\end{figure}

\subsubsection{Spin Seebeck switch.} 
Tuning chemical potentials of two leads can also render us flexible control of thermal spin transport, as displayed in Fig.~\ref{fig3}. When lifting chemical potentials but still below the band bottom of the minority electron spin, we see the spin Seebeck transport is significantly enhanced 
, acting as a {\it spin Seebeck switch}. Physically, this is because lifting chemical potentials increases the coexistence of two electron spins so that $C_v$ increases. When $\mu_L\neq\mu_R$, we also have the spin Seebeck diode, which results from the different temperature responses of $C_L$ and $C_R$ when they have differential chemical potentials in Eq.~(\ref{eq:C}). As we noted, the chemical potential difference, although it acts as a control factor of spin Seebeck effect, can not generate the thermal spin transport [see $I_s=0$ at $\Delta{T}=0$ in Fig.~\ref{fig3} despite the fact that $\mu_L-\mu_R\neq0$]. In insulating-magnetic-molecular junctions, the pure spin transport (spin voltage and spin current) is only conjugated with the thermal transport (temperature bias and heat current). 

When chemical potentials are much above the bottoms of both spin-resolved electron dispersions, the leads behave as good metals and the DOS can be treated as a constant $\rho^0_{v\sigma}$. In this way, Eq.~(\ref{eq:C}) reduces to a temperature-independent coefficient $C_v(\varepsilon_m)=\varepsilon_m\rho^0_{v\uparrow}\rho^0_{v\downarrow}$ (for the spin-half case, $\varepsilon_m=\Omega_0$). As such, we can no longer have the negative differential spin Seebeck effect, for which the temperature-dependent $C_v$ is crucial. Nevertheless, we can achieve the spin Seebeck diode if $C_L{\neq}C_R$ when two leads have different DOS. Even if $C_L=C_R$, we can still build asymmetric system-lead couplings $J_L{\neq}J_R$ so that $I_s$ is asymmetric under temperature interchange $T_L{\leftrightarrow}T_R$ and the rectifying action is retained [see Eq.~(\ref{eq:TLflux})]. 

{ 
\section{Discussions}\label{discussion}
We would like to clarify that our system, which takes a single effective spin as the insulating part in between the leads, can be regarded as
a minimal phenomenological model to mimic the sandwich setup meal/insulating magnet/metal for the spin Seebeck transport at nanoscale. This  macrospin picture is reasoned by the fact that for the coupled spin chain (or cluster, network) in nanoscale ferromagnets with a single-domain state, spins are tightly coupled and thus form an effective coarse-grained macrospin~\cite{macrospin}. 

Microscopically, the spin chain model with exchange interactions can be derived from the tight-binding electron chain model with strong Coulomb interaction. This is achieved by using the Schrieffer-Wolff transformation~\cite{Phillips}, which will naturally give the exchange coupling form Eq.~(\ref{eq:coupling2}) at boundaries. 

\subsection{The double-site case}
For the two-site system, the Hamiltonian after Schrieffer-Wolff transformation is expressed as the same as Eq.~(\ref{eq:H}), except for the new central Hamiltonian with two coupled spin-$1/2$ impurities:
\begin{equation}
H_s=-J \bm S_1 \cdot \bm S_2+\omega_0 S^z_1+\omega_0 S^z_2.
\end{equation}
This central two spin system has four eigenstates, $\textcircled{1}$: $|\uparrow\uparrow\rangle$; $\textcircled{2}$: $|\downarrow\downarrow\rangle$; $\textcircled{3}$: $\frac{1}{\sqrt{2}}(|\uparrow\downarrow\rangle+|\downarrow\uparrow\rangle)$ and $\textcircled{4}$: $\frac{1}{\sqrt{2}}(|\uparrow\downarrow\rangle-|\downarrow\uparrow\rangle)$, with eigenvalues $-{J}/{4}+\omega_0$, $-J/4-\omega_0$, $-J/4$ and $3J/4$, respectively.

The left spin 1 is coupled with the left lead through
$V_{L}=J_L\sum_{k,k'\in{L}}[S_1^{z}(c^{\dag}_{k\uparrow}c_{k'\uparrow}-c^{\dag}_{k\downarrow}c_{k'\downarrow})+S_1^{-}c^{\dag}_{k\uparrow}c_{k'\downarrow}+S_1^{+}c^{\dag}_{k\downarrow}c_{k'\uparrow}]$, which assists the spin state transitions $\textcircled{1}\leftrightarrow\textcircled{3}$, $\textcircled{1}\leftrightarrow\textcircled{4}$, $\textcircled{2}\leftrightarrow\textcircled{3}$, $\textcircled{2}\leftrightarrow\textcircled{4}$. The right spin 2 is coupled with the right lead through $V_{R}=J_R\sum_{k,k'\in{R}}[S_2^{z}(c^{\dag}_{k\uparrow}c_{k'\uparrow}-c^{\dag}_{k\downarrow}c_{k'\downarrow})+S_2^{-}c^{\dag}_{k\uparrow}c_{k'\downarrow}+S_2^{+}c^{\dag}_{k\downarrow}c_{k'\uparrow}]$, which assists the same transitions $\textcircled{1}\leftrightarrow\textcircled{3}$, $\textcircled{1}\leftrightarrow\textcircled{4}$, $\textcircled{2}\leftrightarrow\textcircled{3}$, $\textcircled{2}\leftrightarrow\textcircled{4}$. The transition $\textcircled{1}\leftrightarrow\textcircled{2}$ is not allowed since in the sequential tunneling regime the lead can only flip spins of the central system one by one.
This sequential dynamics is dynamically equivalent to having a phenomenological spin with a finite anisotropy $D$ in contact with two separate electronic baths, without electron transfer across the system.

From the eigen-levels, we know that when inner spin coupling is large, we can effectively have three states, $|\uparrow\uparrow\rangle$, $|\downarrow\downarrow\rangle$ and $\frac{1}{\sqrt{2}}(|\uparrow\downarrow\rangle+|\downarrow\uparrow\rangle)$. The fourth state will be difficult to access by the bath excitation due to the large energy gap. As such, the dynamics of the spin Seebeck transport will then be similar to that across an effective spin 1 of finite anisotropy, with excitation and relaxation by two separate electronic baths.

\subsection{The single-site case}

However,  attention should be paid to the special example -- the single level Anderson impurity model~\cite{Phillips}: 
The central impurity electrons with local Coulomb interaction are hybridized with electrons in the leads. 
In the limit of large Coulomb interaction between electrons of opposite spins on the central level,  the Schrieffer-Wolff transformation reduces the model into a similar geometry setup~\cite{Phillips} as described in Eq.~(\ref{eq:H}) with $S=1/2$ and $D=0$. 

As such, there are in principle three exchange coupling terms~\cite{note}. Among them, two terms are the local exchange coupling of the impurity spin to conduction electron spin density in each lead individually, i.e., $V_L$ and $V_R$ [see Eq.~(\ref{eq:coupling})]; 
While the third term is the coupling of the impurity spin to the tunneling electron spin, which is of the form: 
\begin{equation}
V_{LR}=\sqrt{J_LJ_R}\sum_{\sigma\sigma'}(c^{\dag}_{L\sigma}\bm{\tau}_{\sigma\sigma'}c_{R\sigma'}+c^{\dag}_{R\sigma}\bm{\tau}_{\sigma\sigma'}c_{L\sigma'})\cdot\bm{S},
\label{eq:VLR}
\end{equation}
This third term carries not only the spin current, but also the electric current. Therefore, the existence of this contribution in the special single level Anderson model will remove the infinite property of $ZT$ since Eq.~(\ref{eq:VLR}) brings finite thermal conductivity due to the additional electron transfer. Nevertheless, the properties of ASSE and NDSSE can be still preserved. 

More precisely, similar to obtaining Eq.~(\ref{eq:rate1}), electron tunneling terms in $V_{LR}$  with spin flipping will contribute additional transition rates:
\begin{align}
k^{v\bar{v}}_{01}&=\frac{2\pi J_vJ_{\bar{v}}}{\hbar}\int^{\infty}_{-\infty}d\epsilon\rho_{v\uparrow}(\epsilon+\Omega_0)\rho_{\bar{v}\downarrow}(\epsilon) f^{+}_{v\uparrow}(\epsilon+\Omega_0)f^{-}_{\bar{v}\downarrow}(\epsilon), 
\label{eq:add1}
\\
k^{v\bar{v}}_{10}&=\frac{2\pi J_vJ_{\bar{v}}}{\hbar}\int^{\infty}_{-\infty}d\epsilon\rho_{v\downarrow}(\epsilon)\rho_{\bar{v}\uparrow}(\epsilon+\Omega_0) f^{+}_{v\downarrow}(\epsilon)f^{-}_{\bar{v}\uparrow}(\epsilon+\Omega_0).
\label{eq:add2}
\end{align}
As illustrated in Fig.~\ref{fig4},
$k^{v\bar{v}}_{01}$ describes the rate of the physical process that the local central spin flips from the down state to the up state, and meanwhile  a spin-up electron tunnels from the lead $v$ to a spin-down electron state in the other lead $\bar{v}$ with releasing energy $\Omega_0$ to the flipping of the local central spin; $k^{v\bar{v}}_{10}$ describes the rate of the similar physical process that the local central spin flips from the up state to the down state, and meanwhile  a spin-down electron tunnels from the lead $v$ to a spin-up electron state in the other lead $\bar{v}$ with absorbing energy $\Omega_0$ from the flipping of the local central spin.
\begin{figure}
\vspace{2mm}
\scalebox{0.4}[0.4]{\includegraphics{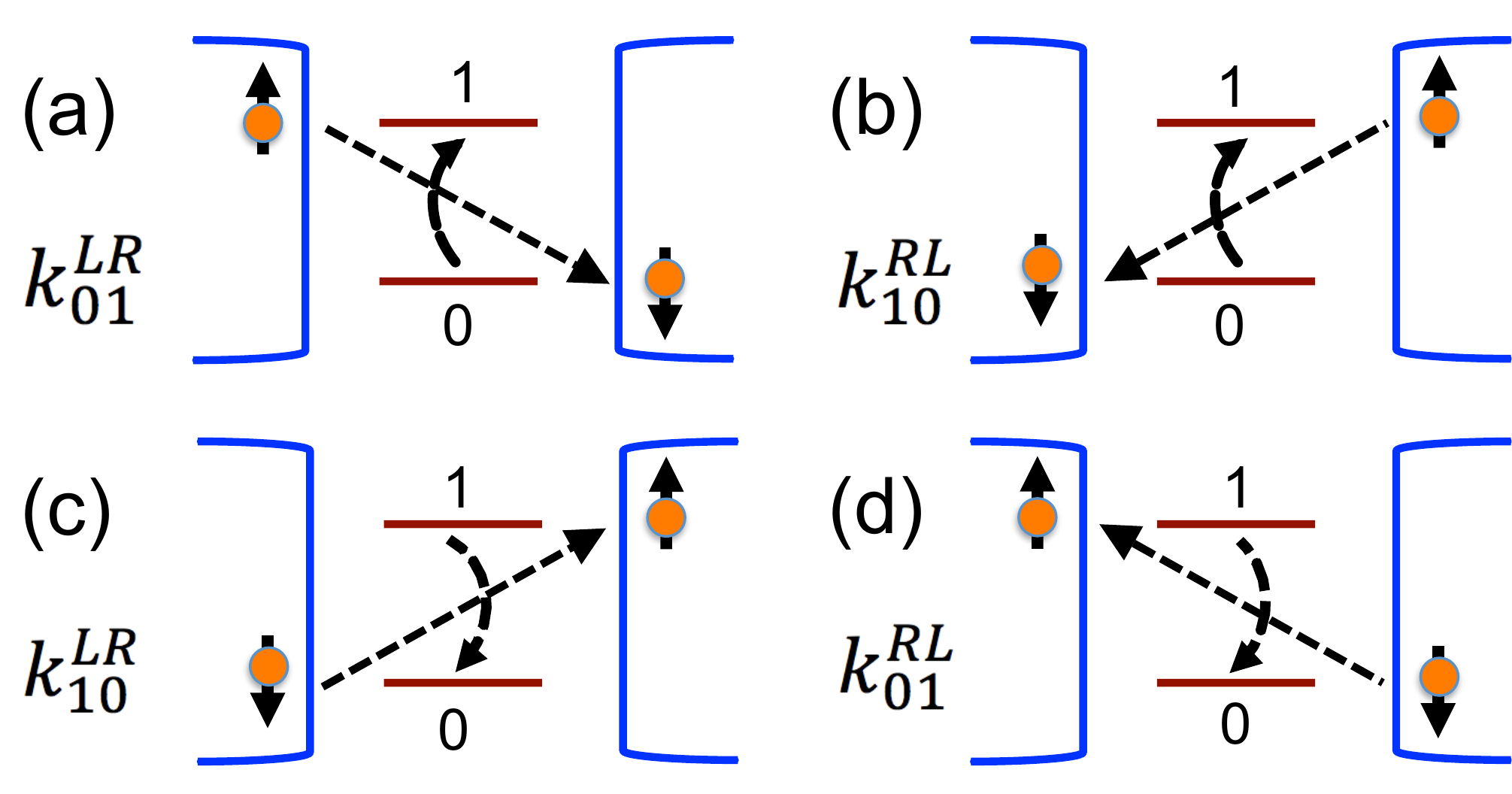}}
\vspace{-5mm}   
\caption{(color online). Schematic illustration of the four physical processes involved in the electron tunneling with spin flipping. } 
\label{fig4}
\end{figure}

As such, following similar procedures as in the main text for the sequential dynamics, the new spin (up) current including additional contributions Eqs.~(\ref{eq:add1}) and (\ref{eq:add2}) is obtained as:
\begin{align}
I^1_s&=\frac{k^L_{01}k^R_{10}-k^R_{01}k^L_{10}}{K}    \nonumber\\
&+\frac{\frac{k^{LR}_{10}+k^{RL}_{10}}{2}(k^L_{01}-k^R_{01})+\frac{k^{LR}_{01}+k^{RL}_{01}}{2}(k^R_{10}-k^L_{10})}{K}.
\label{eq:Is1}
\end{align} 
The first term is reminiscent of Eq.~(\ref{eq:TLflux}), except for the denominator $K=k^L_{01}+k^L_{10}+k^R_{10}+k^R_{01}+k^{LR}_{10}+k^{LR}_{01}+k^{RL}_{01}+k^{RL}_{10}$. The second term mainly describes the contribution from electron tunnelings with spin flipping.
Additionally, the electron tunneling terms in $V_{LR}$ without spin flipping will also contribute to the thermal spin current, which, following the tunneling theory driven by temperature bias (for example, see Ref.~\cite{RenSSE2}), is obtained as
\begin{align}
I^2_s=\frac{2\pi J_LJ_R\langle S_z\rangle^2}{\hbar}\int^{\infty}_{-\infty}&d\epsilon[\rho_{L\uparrow}(\epsilon)\rho_{R\uparrow}(\epsilon)-\rho_{L\downarrow}(\epsilon)\rho_{R\downarrow}(\epsilon)]   \nonumber\\
&\times[f_L(\epsilon)-f_R(\epsilon)].
\end{align}

Therefore, the total spin current is $I_s=I^1_s+I^2_s$. For the same conditions as for the case of up-down lead configuration in Fig.~\ref{fig2}, the spin Seebeck effect is plotted in Fig.~\ref{fig5}(a), from which we see that the thermal spin current profile is clearly modified by the electron tunneling contribution. The ASSE is preserved, although the NDSSE does not occur for this case. In fact, as implied by earlier discussions~\cite{RenSSE1,RenSSE2}, the ASSE is robust once the left and right part are asymmetric while the NDSSE will be more sensitive to the DOS. For the case of choosing Lorentzian type DOS $\rho_{v\uparrow}(\epsilon)=\frac{1}{\pi}\frac{\Gamma}{(\epsilon-\epsilon_{v\uparrow})^2+\Gamma^2}, \rho_{v\downarrow}(\epsilon)=\frac{1}{\pi}\frac{\Gamma}{(\epsilon-\epsilon_{v\downarrow})^2+\Gamma^2}$ for the lead $v$, the thermal spin current profile is plotted in Fig.~\ref{fig5}(b). It shows that although with quantitative changes, the properties of ASSE and NDSSE persist. 

From above discussions, we see that the electron-tunneling contribution in the single impurity Anderson model (a Kondo-type local spin model) will play an important role, which however will disappear in the coupled spin chain and network system. We have considered a phenomenological macrospin model to mimic the coupled spin cluster in the insulating magnetic junctions, it would be interesting in the future to study the coupled microscopic spin model as in Ref.~\cite{Juzar} for the nanoscale spin Seebeck transport.
}

\begin{figure}
\vspace{2mm}
\scalebox{0.4}[0.38]{\includegraphics{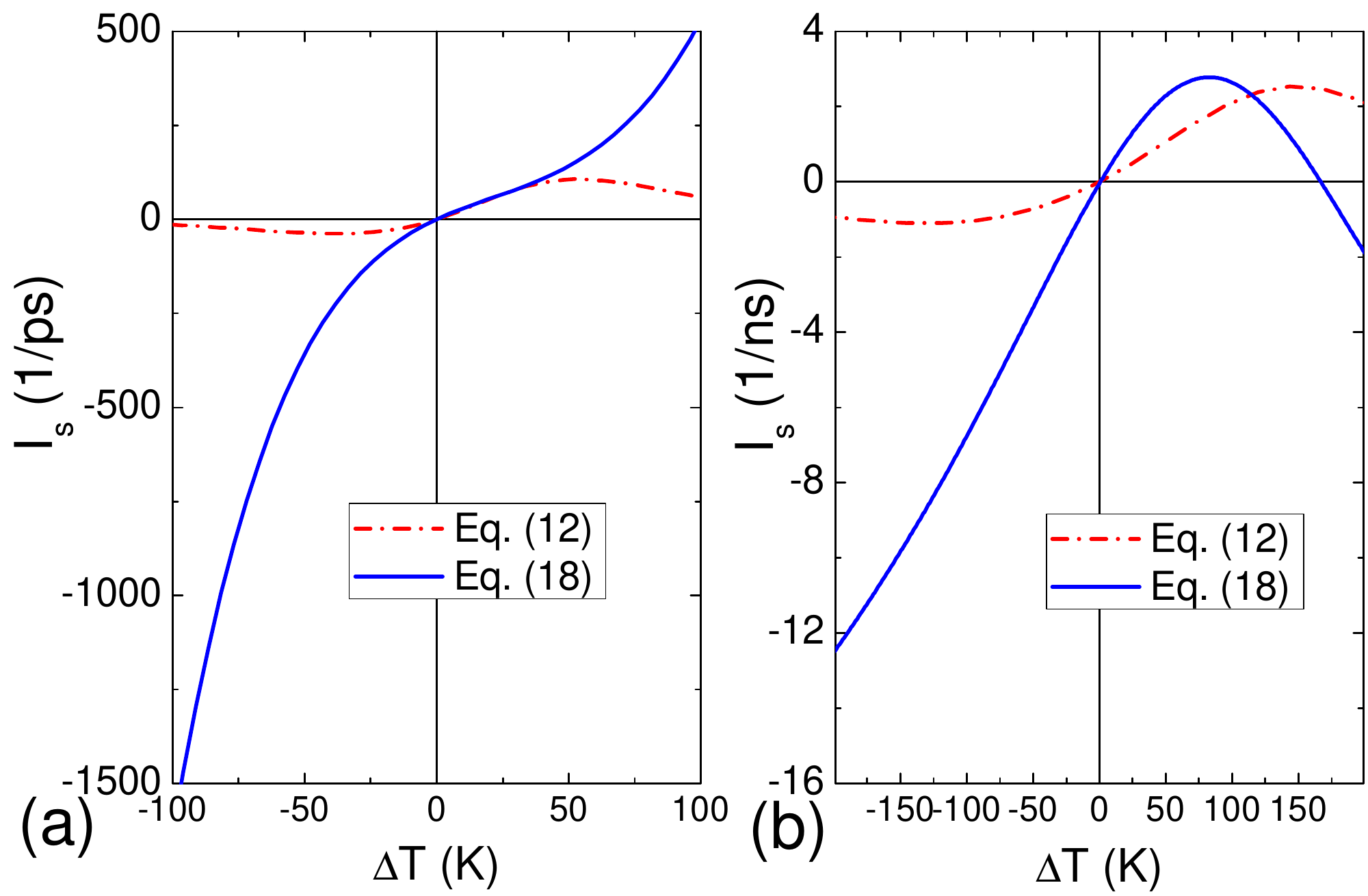}}
\vspace{-3mm}   
\caption{(color online). Rectifying spin Seebeck current with electron tunneling. (a) The case of up-down lead configuration in Fig.~\ref{fig2} with additional electron tunneling contribution. (b) The case of Lorentzian type DOS, with $\epsilon_{L\uparrow}=0, \epsilon_{L\downarrow}=30$ meV for the left lead and $\epsilon_{R\uparrow}=30$ meV, $\epsilon_{R\downarrow}=0$ for the right one.
In both cases, $\rho_{L\uparrow}(\epsilon)\rho_{R\uparrow}(\epsilon)=\rho_{L\downarrow}(\epsilon)\rho_{R\downarrow}(\epsilon)$, so $I^2_s=0$ and the spin current is only contributed by Eq.~(\ref{eq:Is1}). $\Gamma=10$ meV. Other parameters are the same as in Fig.~\ref{fig2}.} 
\label{fig5}
\end{figure}

\section{Conclusion}
In summary, we have studied the nonequilibrium spin Seebeck transport across a charge insulating magnetic junction with localized effective spin. The conjugate-converted thermal-spin transport is assisted by the exchange interactions between the effective macrospin in the center and electrons in metallic leads. We have shown that in contrast with bulk spin Seebeck effect, the figure of merit of the thermal-spin conversion in such nanoscale spin caloritronic devices can be infinite, leading to the ideal Carnot efficiency in the linear response. We have further unravelled the ASSE and NDSSE in the model device,  suggesting that the nanoscale thermal spin rectifier could act as a spin Seebeck diode, spin Seebeck transistor and spin Seebeck switch. {  Cases with electron tunneling are also discussed.} These properties could have various implications in flexible thermal \cite{phononics, Ren2013PRB87} and spin information control \cite{BauerReview,spintronics,magnonics}. It would be desirable in the future to use first-principles approaches to real molecular magnet systems for more realistic calculations.

\begin{acknowledgments}
{This work was carried out under the auspices of  the National Nuclear Security Administration of the U.S. Department of Energy (DOE) at Los Alamos National Laboratory (LANL) under Contract No. DE-AC52-06NA25396, and supported by the LANL Laboratory Directed Research Development Program (J.R.), and  the Swedish Research Council and Wenner-Gren Foundation are also acknowledged (J.F.). The work was supported, in part, by  the Center for Integrated Nanotechnologies, a U.S. DOE Office of Basic Energy Sciences user facility (J.-X.Z.). J.R. thanks J. Thingna for bringing Ref.~\cite{Juzar} to our attention.}
\end{acknowledgments}


\begin{thebibliography}{300}
\bibitem{note} Strictly speaking, after the transformation additional direct electron tunneling terms will also be generated, but still can be removed by tuning the Coulomb interaction strength and the energy level of the impurity.
\end{thebibliography}
\end{document}